# Which optical processes are suitable to make probabilistic single photon sources for quantum cryptography?

Amit Verma and Anirban Pathak
amit.verma@jiit.ac.in, anirbanpathak@yahoo.co.in
Department of Physics and Material Science,
Jaypee Institute of Information Technology University,
A-10, Sector-62, Noida, UP-201307, INDIA.

**ABSTRACT**

Single photon sources to be used in quantum cryptography must show higher order antibunching (HOA). HOA is reported by us in several many wave mixing processes. In the present work we have investigated the possibility of observing HOA in multiwave mixing processes in general. The generalized Hamiltonian is solved for several particular cases in Heisenberg picture and possibility of observing HOA is investigated with the help of criterion of Pathak and Garcia. The generalized interaction Hamiltonian of a multiwave mixing process is considered as ($a^{\dagger\, l}\, b^m\, c^n$ + Hermitian conjugate) where '$a$', '$b$' and '$c$' are the annihilation operators corresponding to pump, Stokes and Signal modes respectively. Several particular cases of the generalized Hamiltonian are solved with the help of short time approximation technique and HOA is reported for pump modes of different multiwave mixing processes. It is also found that HOA can not be observed for the signal and stokes modes in of the cases studied here.

Keywords: Single photon source, Cryptography, Antibunching, Optical processes.

## 1. Introduction

A single photon source (SPS) is very important for quantum computation. In particular, it is essential for secured quantum cryptography. But there is no perfect SPS in reality. Therefore, probabilistic SPS, where probability of simultaneous emission of two, three, four and more photon is less than the emission of a single photon are used. In the well-known antibunched state the rate of simultaneous emission of two photon is less than that of single photon. But the requirement of quantum cryptography is a many photon version of the antibunched state or the higher order antibunched state. Higher order extension of the nonclassical effects (antibunching and squeezing) has been introduced in recent past [1-4]. Among these higher order nonclassical effects higher order squeezing is studied in detail [1,2,5,6] but higher order antibunching which is required for a infinitely secured communication protocol [7] (BB84 protocol) is not yet studied rigorously. The idea of HOA was introduced by Lee in a pioneering paper [3] in 1990, since then it has been predicted in two photon coherent state [4], shadowed negative binomial state [9], trio coherent state [10] and in the interaction of intense laser beam with an inversion symmetric third order nonlinear medium[8]. Recently we have reported a simplified mathematical criterion for observing higher order antibunching (HOA) [8] and with the help of that criterion shown that the HOA is not a rare phenomenon [11] and it can be seen in simple optical processes. But still we don't know what kind of interaction produces HOA? How the signature of HOA is affected by the power of modes in the interaction term of a Hamiltonian for an optical process? The aim of the present work is to provide answer to these questions. To generalize our result we have chosen a new six wave mixing process; explore the possibility of HOA in all modes and our earlier reported results. We have also studied various other multiwave mixing processes and tried to conclude from these set of observations. In the next section we briefly present mathematical criterion for HOA. In section 3 we present a second order operator solution of the equation of motion of six wave mixing process correspond to all possible modes and use that to show the existence of HOA in six wave mixing. Section 4 includes all new observation with earlier reports and we generalize the possibility of HOA to make a probabilistic SPS. Finally section 5 is dedicated to conclusions.

## 2. Mathematical criterion for higher order antibunching:

The criterion of HOA is expressed in terms of higher order factorial moments of number operator. There exist several criterion for the same which are essentially equivalent. Initially, using the negativity of P function [12], Lee introduced the criterion for HOA as

$$R(l,m) = \frac{\langle N_x^{(l+1)} \rangle \langle N_x^{(m-1)} \rangle}{\langle N_x^{(l)} \rangle \langle N_x^{(m)} \rangle} - 1 < 0 \ldots\ldots\ldots \qquad \ldots\ldots\ldots\ldots \quad (1).$$

where N is the usual number operator, $\langle N^{(i)} \rangle = \langle N(N-1)\ldots(N-i+1) \rangle$ is the i$^{th}$ factorial moment of number operator, $\langle\ \rangle$ denotes the quantum average, l and m are integers satisfying the conditions $l \leq m \leq 1$ and the subscript *x* denotes a particular mode. Ba An [10] choose m=1 and reduced the criterion of $l^{th}$ order antibunching to

$$A_{x,l} = \frac{\langle N_x^{(l+1)} \rangle}{\langle N_x^{(l)} \rangle \langle N_x \rangle} - 1 < 0 \ldots\ldots \qquad \ldots\ldots\ldots \qquad \ldots\ldots\ldots (2)$$

or

$$\langle N_x^{(l+1)} \rangle < \langle N_x^{(l)} \rangle \langle N_x \rangle \qquad \ldots\ldots\ldots \qquad \ldots\ldots\ldots \qquad \ldots\ldots\ldots\ldots (3)$$

One can further simplify (3) and obtain the condition for *l*-th order antibunching as

$$d(l) = \langle N_x^{(l+1)} \rangle - \langle N_x \rangle^{l+1} < 0 \qquad \ldots\ldots\ldots \ldots\ldots \qquad \ldots\ (4)$$

This is the criterion of HOA derived independently by Pathak and Garcia [8] and Erenso, Vyas and Singh [13]. Here we can note that d(*l*)=0 and d(*l*)>0 corresponds to higher order coherence and higher order bunching (many photon bunching) respectively. Criterion (4) is exactly the characteristic that is required in a probabilistic single photon source used in quantum cryptography. In other words all the probabilistic single photon sources used in quantum cryptography should satisfy the criteria (4) of HOA.

## 3. Six wave mixing process:

Six wave mixing may happen in different ways. To generalize the possibility of HOA in different ways we have chosen such a way that three photon of frequency $\omega_1$ are absorbed (as pump photon) and two photon of frequency $\omega_2$ and another of frequency $\omega_3$ are emitted. The Hamiltonian representing this particular six wave mixing process is

$$H = a^\dagger a\, \omega_1 + b^\dagger b\, \omega_2 + c^\dagger c\, \omega_3 + g\, (a^{\dagger 3} b^2 c + a^3 b^{\dagger 2} c^\dagger) \quad \ldots\ldots \quad \ldots \quad \ldots\ldots(5)$$

where $a^\dagger$ and *a* are creation and annihilation operators in pump mode which satisfy $[a^\dagger, a]=1$, similarly $b^\dagger$, b and $c^\dagger$, c are creation and annihilation operators in stokes mode and signal mode respectively and g is the coupling constant. Substituting $A = ae^{i\omega_1 t}$, $B = be^{i\omega_2 t}$ and $C = ce^{i\omega_3 t}$ we can write the Hamiltonian (5) as

$$H = A^\dagger A \omega_1 + B^\dagger B\, \omega_2 + C^\dagger C\, \omega_3 + g\, (A^{\dagger 3} B^2 C + A^3 B^{\dagger 2} C^\dagger) \quad \ldots \quad \ldots\ldots \qquad (6)$$

**3.1 Time evolution of A:**

Since the Hamiltonian is known, we can use Heisenberg's equation of motion (with $\hbar = 1$):

$$\dot{A} = \frac{\partial A}{\partial t} + i[H, A] \qquad \ldots\ldots \quad \ldots\ldots \quad \ldots\ldots \quad (7)$$

and short time approximation method [11] to find out the time evolution of the essential operators. From equation (6) we have

$$[H, A] = -A\omega_1 - 3gA^{\dagger 2}B^2 C \qquad \ldots\ldots \quad \ldots\ldots \quad (8)$$

From (7) and (8) we have

$$\dot{A} = iA\omega_1 - iA\omega_1 - i3gA^{\dagger 2}B^2C = -i3gA^{\dagger 2}B^2C \qquad \ldots \quad (9)$$

We can find the second order differential of A using (7) and (9) as

$$\begin{aligned}\ddot{A} &= \frac{\partial \dot{A}}{\partial t} + i[H, \dot{A}] \\ &= 3g^2[6A^{\dagger}A^2 B^{\dagger 2}B^2 C^{\dagger}C + 6AB^{\dagger 2}B^2 C^{\dagger}C - 4A^{\dagger 2}A^3 B^{\dagger}BC^{\dagger}C \\ &\quad - 2A^{\dagger 2}A^3 C^{\dagger}C - A^{\dagger 2}A^3 B^{\dagger 2}B^2 - 4A^{\dagger 2}A^3 B^{\dagger}B - 2A^{\dagger 2}A^3]\end{aligned} \qquad \ldots\ldots \quad (10)$$

Now by using Taylor's series expansion,

$$f(t) = f(0) + t\left(\frac{\partial f(t)}{\partial t}\right)_{t=0} + \frac{t^2}{2!}\left(\frac{\partial^2 f(t)}{\partial t^2}\right)_{t=0} + \ldots\ldots\ldots\ldots \qquad \ldots \quad \ldots\ldots \quad (11)$$

and substituting (9) and (10) in (11) we get

$$\begin{aligned}A(t) &= A - 3igt\, A^{\dagger 2}B^2 C + \frac{3}{2}g^2 t^2[6A^{\dagger}A^2 B^{\dagger 2}B^2 C^{\dagger}C + 6AB^{\dagger 2}B^2 C^{\dagger}C \\ &\quad - 4A^{\dagger 2}A^3 B^{\dagger}BC^{\dagger}C - 2A^{\dagger 2}A^3 C^{\dagger}C - A^{\dagger 2}A^3 B^{\dagger 2}B^2 - 4A^{\dagger 2}A^3 B^{\dagger}B \\ &\quad - 2A^{\dagger 2}A^3]\end{aligned} \qquad \ldots \quad (12)$$

The Taylor series is valid when t is small, so this solution is valid for a short time and that is why it is called short time approximation. This is a very strong technique since this straight forward prescription is valid for any optical process where interaction time is short. After obtaining the analytic expression for time evolution of annihilation operator, now we can use it to check whether it satisfies condition (4) or not.

Let us start with the possibility of observing first order antibunching in pump mode. From equation (12), Hermitian conjugate of A(t) can be written as

$$\begin{aligned}A^{\dagger}(t) &= A^{\dagger} + 3igt\, A^2 B^{\dagger 2}C^{\dagger} + \frac{3}{2}g^2 t^2[6A^{\dagger 2}AB^{\dagger 2}B^2 C^{\dagger}C + 6A^{\dagger}B^{\dagger 2}B^2 C^{\dagger}C - 4A^{\dagger 3}A^2 B^{\dagger}BC^{\dagger}C \\ &\quad - 2A^{\dagger 3}A^2 C^{\dagger}C - A^{\dagger 3}A^2 B^{\dagger 2}B^2 - 4A^{\dagger 3}A^2 B^{\dagger}B - 2A^{\dagger 3}A^2]\end{aligned} \qquad \ldots \quad (13)$$

Hence using (12) and (13), number operator $N_A(t)$ can be derived as

$$\begin{aligned}N_A(t) &= A^{\dagger}(t)A(t) \\ &= A^{\dagger}A - 3igt(A^{\dagger 3}B^2 C - A^3 B^{\dagger 2}C^{\dagger}) \\ &\quad + 3g^2 t^2[9A^{\dagger 2}A^2 B^{\dagger 2}B^2 C^{\dagger}C + 18A^{\dagger}AB^{\dagger 2}B^2 C^{\dagger}C + 6B^{\dagger 2}B^2 C^{\dagger}C \\ &\quad - 4A^{\dagger 3}A^3 B^{\dagger}BC^{\dagger}C - 4A^{\dagger 3}A^3 B^{\dagger}B - 2A^{\dagger 3}A^3 C^{\dagger}C \\ &\quad - A^{\dagger 3}A^3 B^{\dagger 2}B^2 - 2A^{\dagger 3}A^3]\end{aligned} \qquad \ldots\ldots \quad (14)$$

Taking expectation value of $N_A(t)$ with respect to $|\alpha\rangle|0\rangle|0\rangle$, we can get

$$\langle N_A(t)\rangle_{\alpha} = |\alpha|^2 - 6g^2 t^2 |\alpha|^6 \qquad \ldots\ldots \quad \ldots\ldots \quad (15)$$

where $A|\alpha\rangle = \alpha|\alpha\rangle$. By using this straight forward description we can get

$$\langle N_A^{(2)}(t)\rangle_{\alpha} = \langle A^{\dagger 2}(t)A^2(t)\rangle = |\alpha|^4 - 12g^2 t^2[|\alpha|^6 + |\alpha|^8] \qquad \ldots\ldots \quad \ldots\ldots \quad (16)$$

and

$$\left\langle N_A^{(3)}(t)\right\rangle_\alpha = \left\langle A^{\dagger 3}(t)A^3(t)\right\rangle = |\alpha|^6 - 6g^2t^2[2|\alpha|^6 + 6|\alpha|^8 + 3|\alpha|^{10}] \qquad \ldots\ldots \qquad \ldots\ldots\ldots (17)$$

Now by using (15) and (16-17) we can show that the pump mode for six wave mixing process satisfies the criterion of normal antibunching (4) and second order antibunching. Since,

$$\begin{aligned}
d_A(1) &= \left\langle N_A^{(2)}(t)\right\rangle_\alpha - \left\langle N_A(t)\right\rangle_\alpha^2 \\
&= |\alpha|^4 - 12g^2t^2[|\alpha|^6 + |\alpha|^8] - |\alpha|^4 + 12g^2t^2|\alpha|^8 \qquad \ldots\ldots\ldots \ldots\ldots\ldots (18)\\
&= -12g^2t^2|\alpha|^6
\end{aligned}$$

and

$$\begin{aligned}
d_A(2) &= \left\langle N_A^{(3)}(t)\right\rangle_\alpha - \left\langle N_A(t)\right\rangle_\alpha^3 \\
&= |\alpha|^6 - 6g^2t^2[2|\alpha|^6 + 6|\alpha|^8 + 3|\alpha|^{10}] - |\alpha|^6 + 18g^2t^2|\alpha|^{10} \qquad \ldots\ldots\ldots (19)\\
&= -12g^2t^2[|\alpha|^6 + 3|\alpha|^8]
\end{aligned}$$

is always negative. Hence normal antibunching and higher order antibunching (second order) exists in pump mode. Expectation value are taken over |α>|0>|0> physically means that initially a coherent state (Laser) is used as pump and before the interaction of the pump with atom, there was no photon in signal mode (b) or stokes mode (c). Thus the pump interacts with the atom and causes excitation followed by emission.

Now our aim for studying this particular problem is whether stokes mode (expectation value over |0>|β>|0>) and signal mode (expectation value over |0>|0>|γ>) also shows antibunching or not for time evolution of B and C in six wave mixing process.

**3.2 Time evolution of B:**
Following same procedure as mentioned in subsection 3.1, we can get

$$\dot{B} = -2igA^3B^\dagger C^\dagger \qquad \ldots\ldots \qquad \ldots\ldots (20)$$

and

$$\ddot{B} = 2g^2[A^{\dagger 3}A^3B^\dagger B^2 + 2A^{\dagger 3}A^3BC^\dagger C + 2A^{\dagger 3}A^3B - 9A^{\dagger 2}A^2B^\dagger B^2C^\dagger C \qquad (21)$$
$$-18A^\dagger AB^\dagger B^2C^\dagger C - 6B^\dagger B^2C^\dagger C]$$

Hence Time evolution of B can be written as, using Taylor's expansion,

$$\begin{aligned}
B(t) &= B - 2igtA^3B^\dagger C^\dagger \\
&+ g^2t^2[A^{\dagger 3}A^3B^\dagger B^2 + 2A^{\dagger 3}A^3BC^\dagger C + 2A^{\dagger 3}A^3B - 9A^{\dagger 2}A^2B^\dagger B^2C^\dagger C \qquad (22)\\
&- 18A^\dagger AB^\dagger B^2C^\dagger C - 6B^\dagger B^2C^\dagger C]
\end{aligned}$$

and number operator correspond to stokes mode is,

$$\begin{aligned}
N_B(t) &= B^\dagger(t)B(t) \\
&= B^\dagger B - 2igt[A^3B^{\dagger 2}C^\dagger - A^{\dagger 3}B^2C] \\
&+ 2g^2t^2[A^{\dagger 3}A^3B^{\dagger 2}B^2 + 4A^{\dagger 3}A^3B^\dagger BC^\dagger C + 4A^{\dagger 3}A^3B^\dagger B + 2A^{\dagger 3}A^3C^\dagger C \qquad \ldots(23)\\
&- 9A^{\dagger 2}A^2B^{\dagger 2}B^2C^\dagger C - 18A^\dagger AB^{\dagger 2}B^2C^\dagger C - 6B^{\dagger 2}B^2C^\dagger C + 2A^{\dagger 3}A^3]
\end{aligned}$$

Expectation value of $N_B(t)$ with respect to |α>|0>|0> is $\left\langle N_B(t)\right\rangle_\alpha = 4g^2t^2|\alpha|^6$, with respect to |0>|β>|0> is $\left\langle N_B(t)\right\rangle_\beta = |\beta|^2$ and for |0>|0>|γ> is $\left\langle N_B(t)\right\rangle_\gamma = 0$. Similarly we can obtain Expectation values of second factorial moment $N_B^{(2)}(t)$ and third factorial moment $N_B^{(3)}(t)$ of number operator $N_B(t)$ with respect to |α>|0>|0> are $\left\langle N_B^{(2)}(t)\right\rangle_\alpha = 4g^2t^2|\alpha|^6$ and $\left\langle N_B^{(3)}(t)\right\rangle_\alpha = 0$, with respect to |0>|β>|0> are $\left\langle N_B^{(2)}(t)\right\rangle_\beta = |\beta|^4$ and

$\langle N_B^{(3)}(t)\rangle_\beta = |\beta|^6$, finally for |0>|0>|γ> are 0. Therefore normal and second order antibunching for B over pump mode are $d_{B\alpha}(1)= 4g^2t^2|\alpha|^6$ and $d_{B\alpha}(2)= 0$, over stokes mode and signal mode are absent.

### 3.3 Time evolution for C:

Time evolution of C can be derived by using Taylor's expansion,

$$C(t) = C - igtA^3B^{\dagger 2}$$
$$+ \frac{1}{2}g^2t^2[A^{\dagger 3}A^3B^{\dagger 2}B^2C + 4A^{\dagger 3}A^3B^\dagger BC + 2A^{\dagger 3}A^3C - A^{\dagger 3}A^3B^{\dagger 2}B^2 \ldots \quad (24)$$
$$- 9A^{\dagger 2}A^2B^{\dagger 2}B^2 - 18A^\dagger AB^{\dagger 2}B^2 - 6B^{\dagger 2}B^2]$$

and Number operator $N_C(t)$ is,

$$N_C(t) = C^\dagger C - igt[A^3B^{\dagger 2}C^\dagger - A^{\dagger 3}B^2C]$$
$$+ \frac{1}{2}g^2t^2[2A^{\dagger 3}A^3B^{\dagger 2}B^2C^\dagger C + 8A^{\dagger 3}A^3B^\dagger BC^\dagger C + 4A^{\dagger 3}A^3C^\dagger C$$
$$- A^{\dagger 3}A^3B^{\dagger 2}B^2C^\dagger - 9A^{\dagger 2}A^2B^{\dagger 2}B^2C^\dagger - 18A^\dagger AB^{\dagger 2}B^2C^\dagger - 6B^{\dagger 2}B^2C^\dagger \quad \ldots (25)$$
$$- A^{\dagger 3}A^3B^{\dagger 2}B^2C - 9A^{\dagger 2}A^2B^{\dagger 2}B^2C - 18A^\dagger AB^{\dagger 2}B^2C - 6B^{\dagger 2}B^2C]$$

Expectation values of $N_C(t)$ with respect to |α>|0>|0> is $\langle N_C(t)\rangle_\alpha = 0$, with respect to |0>|β>|0> is $\langle N_C(t)\rangle_\beta = 0$ and with respect to |0>|0>|γ> is $\langle N_C(t)\rangle_\gamma = |\gamma|^2$. Similarly we can obtain Expectation values of second factorial moment $N_C^{(2)}(t)$ and third factorial moment $N_C^{(3)}(t)$ of number operator N(t) with respect to |α>|0>|0>, |0>|β>|0> are 0 and finally with respect to |0>|0>|γ> are $\langle N_C^{(2)}(t)\rangle_\gamma = |\gamma|^4$ and $\langle N_C^{(3)}(t)\rangle_\gamma = |\gamma|^6$.

Hence $d_C(1)$ and $d_C(2)$ are always zero for all modes. Therefore neither the normal antibunching nor the second order antibunching is present in mode C. Earlier our group has reported normal and higher order antibunching for pump mode in another kind of six wave mixing [11] whose interaction term in the Hamiltonian was $a^{\dagger 2}b^3c$, In the same manner we have obtained the results for all other modes in six wave mixing ($a^{\dagger 2}b^3c$) also. Results are given in table 1.

| Sr. No | Optical processes | Interaction Term | Expectation value wrt * / d parameter | Mode A | | | Mode B | | | Mode C | | |
|---|---|---|---|---|---|---|---|---|---|---|---|---|
| | | | | \|α> | \|β> | \|γ> | \|α> | \|β> | \|γ> | \|α> | \|β> | \|γ> |
| 1 | Six wave mixing (new) | $A^{\dagger 3}B^2C$ | d(1) | $-12g^2t^2\|\alpha\|^6$ | 0 | 0 | $4g^2t^2\|\alpha\|^6$ | 0 | 0 | 0 | 0 | 0 |
| | | | d(2) | $-12g^2t^2[\|\alpha\|^6+3\|\alpha\|^8]$ | 0 | 0 | 0 | 0 | 0 | 0 | 0 | 0 |
| 2 | Six wave mixing (earlier) | $A^{\dagger 2}B^3C$ | d(1) | $-12g^2t^2\|\alpha\|^4$ | 0 | 0 | $36g^2t^2\|\alpha\|^4$ | 0 | 0 | 0 | 0 | 0 |
| | | | d(2) | $-36g^2t^2\|\alpha\|^6$ | 0 | 0 | 0 | 0 | 0 | 0 | 0 | 0 |

Table1

* Expectation values are taken with respect to |α>|0>|0>, |0>|β>|0> and |0>|0>|γ> respectively for each mode. These states physically correspond to the initial states. Vacuum state is not written in the top of the table (i.e. |α>|0>|0> is written as |α>).

### 4. Other optical processes:

Earlier our group has reported normal and higher order antibunching for pump mode only in other optical processes for example five wave mixing [14], four wave mixing [11] and second harmonic generation [11]. In the present work we have studied normal and higher order antibunching in two types of six wave mixing processes for all modes. Now we want to generalize our results on the basis of earlier and new observations related to HOA in optical processes. All the results are mentioned in table2.

| Sr. No | Optical processes | Interaction Term | Expectation values w r t * d parameter | mode A | | | mode B | | | mode C | | |
|---|---|---|---|---|---|---|---|---|---|---|---|---|
| | | | | $|\alpha\rangle$ | $|\beta\rangle$ | $|\gamma\rangle$ | $|\alpha\rangle$ | $|\beta\rangle$ | $|\gamma\rangle$ | $|\alpha\rangle$ | $|\beta\rangle$ | $|\gamma\rangle$ |
| 1 | Six wave mixing (new) | $A^{\dagger 3}B^2C$ | d(1) | $-12g^2t^2|\alpha|^6$ | 0 | 0 | $4g^2t^2|\alpha|^6$ | 0 | 0 | 0 | 0 | 0 |
| | | | d(2) | $-12g^2t^2[|\alpha|^6+3|\alpha|^8]$ | 0 | 0 | 0 | 0 | 0 | 0 | 0 | 0 |
| 2 | Six wave mixing (earlier) | $A^{\dagger 2}B^3C$ | d(1) | $-12g^2t^2|\alpha|^4$ | 0 | 0 | $36g^2t^2|\alpha|^4$ | 0 | 0 | 0 | 0 | 0 |
| | | | d(2) | $-36g^2t^2|\alpha|^6$ | 0 | 0 | 0 | 0 | 0 | 0 | 0 | 0 |
| 3 | Four wave mixing | $A^{\dagger 2}BC$ | d(1) | $-2g^2t^2|\alpha|^4$ | 0 | 0 | 0 | 0 | 0 | 0 | 0 | 0 |
| | | | d(2) | $-6g^2t^2|\alpha|^6$ | 0 | 0 | 0 | 0 | 0 | 0 | 0 | 0 |
| 4 | Second harmonic generation | $A^{\dagger 2}B$ | d(1) | $-2g^2t^2|\alpha|^4$ | 0 | 0 | 0 | 0 | 0 | 0 | 0 | 0 |
| | | | d(2) | $-6g^2t^2|\alpha|^6$ | 0 | 0 | 0 | 0 | 0 | 0 | 0 | 0 |
| 5 | Five wave mixing | $A^{\dagger 3}B^2$ | d(1) | $-12g^2t^2|\alpha|^6$ | 0 | 0 | 0 | 0 | 0 | 0 | 0 | 0 |
| | | | d(2) | $-12g^2t^2[3|\alpha|^8+|\alpha|^6]$ | 0 | 0 | 0 | 0 | 0 | 0 | 0 | 0 |
| 6 | Third harmonic generation | $A^{\dagger 3}B$ | d(1) | $-6g^2t^2|\alpha|^6$ | 0 | 0 | 0 | 0 | 0 | 0 | 0 | 0 |
| | | | d(2) | $-6g^2t^2(3|\alpha|^8+|\alpha|^6)$ | 0 | 0 | 0 | 0 | 0 | 0 | 0 | 0 |
| 7 | Tri-linear parametric process | $A^{\dagger}BC$ | d(1) | 0 | 0 | 0 | 0 | 0 | 0 | 0 | 0 | 0 |
| | | | d(2) | 0 | 0 | 0 | 0 | 0 | 0 | 0 | 0 | 0 |

Table 2

* Expectation values are taken with respect to $|\alpha\rangle|0\rangle|0\rangle$, $|0\rangle|\beta\rangle|0\rangle$ and $|0\rangle|0\rangle|\gamma\rangle$ respectively for each mode. These states physically correspond to the initial states. Vacuum state is not written in the top of the table (i.e. $|\alpha\rangle|0\rangle|0\rangle$ is written as $|\alpha\rangle$).

## 5 Conclusions:

From table 2, it is clear that except tri-linear parametric process all the physical systems studied show antibunching and HOA in pump mode only. For stokes mode and signal mode antibunching and HOA is absent. When the expectation values are taken over $|\alpha\rangle|0\rangle|0\rangle$, physically that means a choice of a coherent state (Laser) as pump and before the interaction of the pump with atom, there was no photon in signal mode (b) or stokes mode(c). Thus the pump interacts with the atom and causes excitation followed by emission. This state is much more physical compared to the other two states used in the present work. Interestingly all the interesting results appeared for this particular choice of initial state. It is also observed that depth of nonclassicality can be tuned with the help of number of photons present in pump mode. It is also clear that the HOA would not have been observed if we would have considered first order operator solutions (first order in g), on the other hand if we use second order operator solutions then the depth of nonclassicality is found to increase monotonically with the increase of input photon number ($|\alpha|^2$). This monotonic growth may be seized in presence of higher order terms. Following the work of Vyas and Singh [15] one can easily make an experimental setup to observe HOA (see [16]) and verify the prediction of the present work. In such an experiment it would not be wise to choose a large value of ($|\alpha|^2$) as in that case the detector will essentially record a superposition of antibunched and coherent pulse and with increase of ($|\alpha|^2$) the probability of detecting original (non-absorbed) coherent (Poisonian) photon will also increase. Most of the physical systems shown in the present paper are simple and easily achievable in laboratories and thus it opens up the possibility of experimental observation of HOA and will decide suitable probabilistic single photon source.

**Acknowledgement:** A. P. thanks to DST, India for partial financial support through the project grant SR\FTP\PS-13\2004.